\PassOptionsToPackage{hyphens}{url}
\documentclass[sn-standardnature]{sn-jnl}% Standard Nature Portfolio 
\theoremstyle{thmstyleone}%
\theoremstyle{thmstyletwo}%
\theoremstyle{thmstylethree}%
\usepackage{booktabs} %for the three line table.
\usepackage{threeparttable} % for multiple rows footnote in the table
\usepackage{cleveref} %for the multiple reference
\usepackage{graphicx}
\usepackage{hyperref} %for the hyperlink
\hypersetup{
	colorlinks=true,
	linkcolor=blue,
	filecolor=blue,      
	urlcolor=blue,
	citecolor=blue,
}
\usepackage{listings} %for the code.
\usepackage{textcomp}
\usepackage{siunitx}

\usepackage{csvsimple}
\usepackage{rotating}
\usepackage{makecell}
\usepackage{caption}
\usepackage{subcaption}
\usepackage{pifont} % for number with ring
\usepackage{stmaryrd} % for \llbracket and \rrbracket
\begin{document}

\title[ ]{Deep learning based Image Compression for Microscopy Images: An Empirical Study}

\author[1]{\fnm{Yu} \sur{Zhou}}

\author[1,2]{\fnm{Jan} \sur{Sollmann}}

\author*[1]{\fnm{Jianxu} \sur{Chen}}\email{jianxu.chen@isas.de}

\affil[1]{\orgname{Leibniz-Institut für Analytische Wissenschaften – ISAS – e.V.}, \orgaddress{\city{Dortmund}, \country{Germany}}}
\affil[2]{\orgname{Ruhr University Bochum}, \orgaddress{\city{Bochum}, \country{Germany}}}

\abstract{With the fast development of modern microscopes and bioimaging techniques, an unprecedentedly large amount of imaging data are being generated, stored, analyzed, and shared through networks. The size of the data poses great challenges for current data infrastructure. One common way to reduce the data size is by image compression. Effective image compression methods could help reduce the data size significantly without losing necessary information, and therefore reduce the burden on data management infrastructure and permit fast transmission through the network for data sharing or cloud computing. This present study analyzes multiple classic and deep learning based image compression methods, as well as an empirical study on their impact on downstream deep learning based image processing models. We used deep learning based label-free prediction models (i.e., predicting fluorescent images from bright field images) as an example downstream task for the comparison and analysis of the impact of image compression. Different compression techniques are compared in compression ratio, image similarity and, most importantly, the prediction accuracy of label-free models on original and compressed images. We found that AI-based compression techniques largely outperform the classic ones with minimal influence on the downstream 2D label-free tasks. In the end, we hope the present study could shed light on the potential of deep learning based image compression and raise the awareness of potential impacts of image compression on downstream deep learning models for analysis. The codebase has been released at: \url{https://github.com/MMV-Lab/data-compression}}

\maketitle

\section{Introduction}
\label{sec:introduction}
Image compression is the process of reducing the size of digital images while retaining the useful information for reconstruction. This is achieved by removing redundancies in the image data, resulting in a compressed version of the original image that requires less storage space and can be transmitted more efficiently. In many fields of research, including microscopy, high-resolution images are often acquired and processed, leading to significant challenges in terms of storage and computational resources. In particular, researchers in the microscopy image analysis field are often faced with infrastructure limitations, such as limited storage capacity or network bandwidth. Image compression can help mitigate such challenges, allowing researchers to store and transmit images efficiently without compromising their quality and validity. Lossless image compression refers to the compression techniques preserving every bit of information in data and make error-free reconstruction, ideal for applications where data integrity is paramount. However, the limited capability in size reduction, e.g., 2 $\sim$ 3 compression ratio \cite{walker2023comparison} is far from sufficient for easing the data explosion crisis. In this work, we focuses on lossy compression methods, where some information lost may occur but can yield significantly higher compression ratio. Despite lossy compression techniques (both classic and deep learning based) are widely employed in the computer vision field, their feasibility and impact in the field of biological microscopy images remain largely underexplored.

In this paper, we proposed a two-phase evaluation pipeline, compression algorithms comparison and downstream tasks analysis in the context of microscopy images. In order to fully explore the impact of lossy image compression on downstream image analysis tasks, we employed a set of label-free models, a.k.a., in-silico labelling \cite{christiansen2018silico}. Label-free model denotes a deep learning approach capable of directly predicting fluorescent images from transmitted light bright-field images \cite{Ounkomol2018Label-freeMicroscopy}. Considering the large amount bright-field images are being used in regular biological studies, it is of great importance that such data compression techniques can be utilized without compromising the prediction quality. 

Through intensive experiments we demonstrated that deep learning based compression methods can outperform the classic algorithms in terms of compression ratio and post-compression reconstruction quality, and their impact on the downstream label-free task, indicating their huge potentials in the bioimaging field. Meanwhile, we made an preliminary attempt to build 3D compression models and reported the current limitation and possible future directions. Overall, we want to raise the awareness of the importance and potentials of deep learning based compression techniques, and hopefully help a strategical planning of future data infrastructure for bioimaging. 

Specifically, the main contribution of the paper is:
\begin{enumerate}
    \item Benchmark common classic and deep learning-based image compression techniques in the context of 2D grayscale bright-field microscopy images.
    \item Empirically investigate the impact of data compression to the downstream label-free tasks. 
    \item Expand the scope of the current compression analysis for 3D microscopy images.
\end{enumerate}
The remaining of this paper is organized as follows: \cref{sec:related_work} will introduce classic and deep learning-based image compression techniques, followed by the method descriptions in \cref{sec:methodology} and experimental settings in \cref{sec:implementation_detail}. Results and discussions will be presented in \cref{sec:result_and_discussion} with conclusions in \cref{sec:conclusion}.

\section{Related Works}
\label{sec:related_work}
The classic data compression techniques are well studied in the last few decades, with the development of JPEG \cite{Wallace1992jpeg}, a popular lossy compression algorithm since 1992, and its successors JPEG 2000 \cite{Marcellin2002AnJPEG-2000}, JPEG XR \cite{Dufaux2009TheNutshell}, etc. In recent years some more powerful algorithms, such as Limited Error Raster Compression (LERC) are proposed \cite{HowAPI}. Generally, the compression process approximately involves the following steps: color transform (with optional downsampling), Domain transform (e.g., Discrete Cosine Transform (DCT) 
\cite{ahmed1974discrete} in JPEG), quantization and further lossless coding (e.g. run-length encoding (RLE) or Huffmann coding \cite{huffman1952method}).

\begin{figure}[!htbp]
  \centering
  \includegraphics[width=\linewidth]{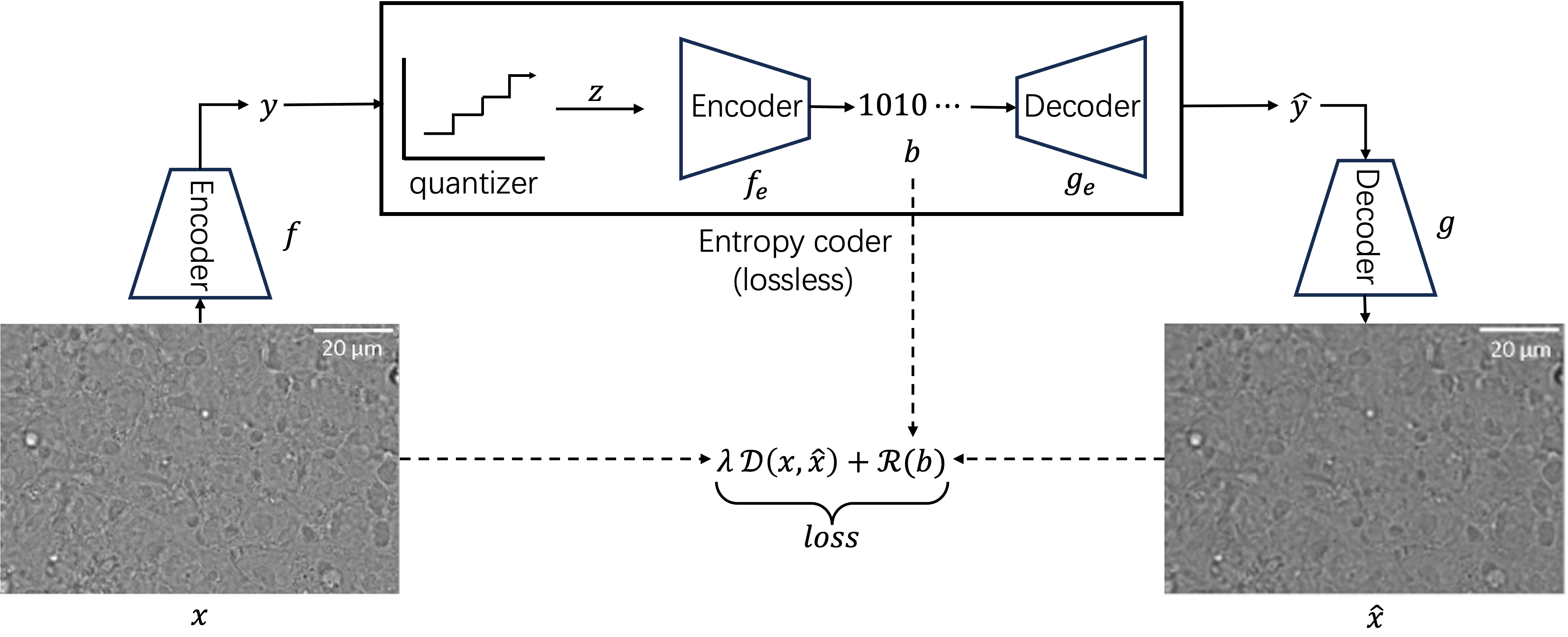}
  \caption{The workflow of a typical learned-based lossy image compression. The raw image $\mathbf{x}$ is fed into the encoder $f$ and obtain the low-dimensional latent representation $\mathbf{y}$. Then the lossless entropy coder can further exploit the information redundency: $\mathbf{y}$ will be firstly quantized to $\mathbf{z} \in \mathbb{Z}^n$, and then compressed to the bitstream $\mathbf{b}$ by the entropy encoder $f_e$. This bitstream can be stored for transmission or further decompression. The corresponding entropy decoder $g_e$ is responsible for the decompression and yield the reconstructed latent representation $\mathbf{\hat{y}}$. Lastly, $\mathbf{\hat{y}}$ is transmitted to the neural decoder $g$, yielding the reconstructed image $\mathbf{\hat{x}}$. The loss function of the system is composed of 2 parts: distortion $\mathcal{D}$ and rate $\mathcal{R}$. Distortion represents the reconstruction quality (e.g. SSIM between $\mathbf{x}$ and $\mathbf{\hat{x}}$) while rate focuses more on the compression ability. $\lambda$ acts as the hyper-parameter to balance the Rate-Distortion trade-off.}
  \label{fig:encoder_decoder}
\end{figure}

Recently, deep learning based image compression gains its popularity thanks to the significantly improved compression performance. Roughly speaking, a deep learning based compression model consists of two sub-networks: a neural encoder $f$ that compresses the image data and a neural decoder $g$ that reconstructs the original image from the compressed representation. Besides, the latent representation will be further losslessly compressed by some entropy coding techniques (e.g. arithmetic coding \cite{Rissanen1979Arithmetic}) as seen in \cref{fig:encoder_decoder}. Specially, the latent vector will be firstly discretized into $\mathbf{z}$: $\mathbf{z} = \llbracket f(\mathbf{X}) \rrbracket, \mathbf{z} \in \mathbb{Z}^n$. Afterwards, $\mathbf{z}$ will be encoded/decoded by the entropy coder ($f_e/g_e$) and decompressed by the neural decoder $g$: $\hat{\mathbf{X}} = g(g_e(f_e(\mathbf{z})))$. The objective is to minimize the loss function containing Rate-Distortion trade-off \cite{cover1999elements,shannon1959coding}:
%--------------------------------------------------------
\begin{gather}
    \mathcal{R}:=\mathbb{E}\left[-\log _2 P(\mathbf{z})\right] \label{eq:R} \\
    \mathcal{D}:=\mathbb{E}[\rho(\mathbf{X}, \hat{\mathbf{X}})] \label{eq:D} \\
    \mathcal{L}:= \mathcal{R} + \lambda \cdot \mathcal{D} \label{eq:L}
\end{gather}
%--------------------------------------------------------
where $\mathcal{R}$ corresponds to the rate loss term, which highlights the compression ability of the system. $P$ is the entropy model that provides prior probability to the entropy coding, and $-\log _2 P(\mathbf{\cdot})$ denotes the information entropy and can approximately estimate the optimal compression ability of the entropy encoder $f_e$, defined by the Shannon theory \cite{shannon1959coding,shannon1948mathematical}. $\mathcal{D}$ is the distortion term, which can control the reconstruction quality. $\rho$ is the norm or perceptual metric, e.g. MSE, MS-SSIM \cite{wang2003multiscale}, etc.  The trade-off between these two terms is achieved by the scale hyper-parameter $\lambda$.

%--------------------------------------------------------
%--------------------------------------------------------
Since the lossless entropy coding entails the accurate modeling of the prior probability of the quantized latent representation $P(\mathbf{z})$, Ball\'{e} et al. \cite{Balle2018VariationalHyperprior} justified that there exists statistical dependencies in the latent representation using current fully factorized entropy model, which will lead to suboptimal performance and not adaptive to all images. To further improve the entropy model, Ball\'{e} et al. proposes a \textit{hyperprior} approach \cite{Balle2018VariationalHyperprior}, where a hyper latent $\mathbf{h}$ (also called \textit{side information}) is generated by the auxillary neural encoder $f_a$ from the latent space $\mathbf{y}$: $\mathbf{h} = f_a(\mathbf{y})$, then the scale parameter of the entropy model can be estimated by the output of the auxillary decoder $g_a$: $\phi = g_a(E_a(\mathbf{h}))$, so that the entropy model can be adaptively adjusted by the input image $\mathbf{x}$, with the bit-rate further enhanced. Minnen et al.  \cite{Minnen2018JointCompression} extended the work to get the more reliable entropy model by jointly combining the data from the above mentioned \textit{hyperprior} and the proposed autoregressive \textit{Context Model}. 

Besides the improvement in the entropy model, lots of effort are also put into the enhancement of the network architecture. Ball\'{e} et al. \cite{balle2015density} replaced the normal RELU activation to the proposed Generalized Division Normalization (GDN) module in order to better capture the image statistics. Johnston et al. \cite{johnston2019computationally} optimized the GDN module in a computational efficient manner without sacrificing the accuracy. Cheng et al. \cite{Cheng2020LearnedModules} introduced the skip connection and attention mechanism. The transformer-based auto-encoder was also reported for data compression in recent years \cite{zhu2021transformer}.

\section{Methodology}
\label{sec:methodology}
\begin{figure}[!tbp]
        \centering
        \includegraphics[width=\linewidth]{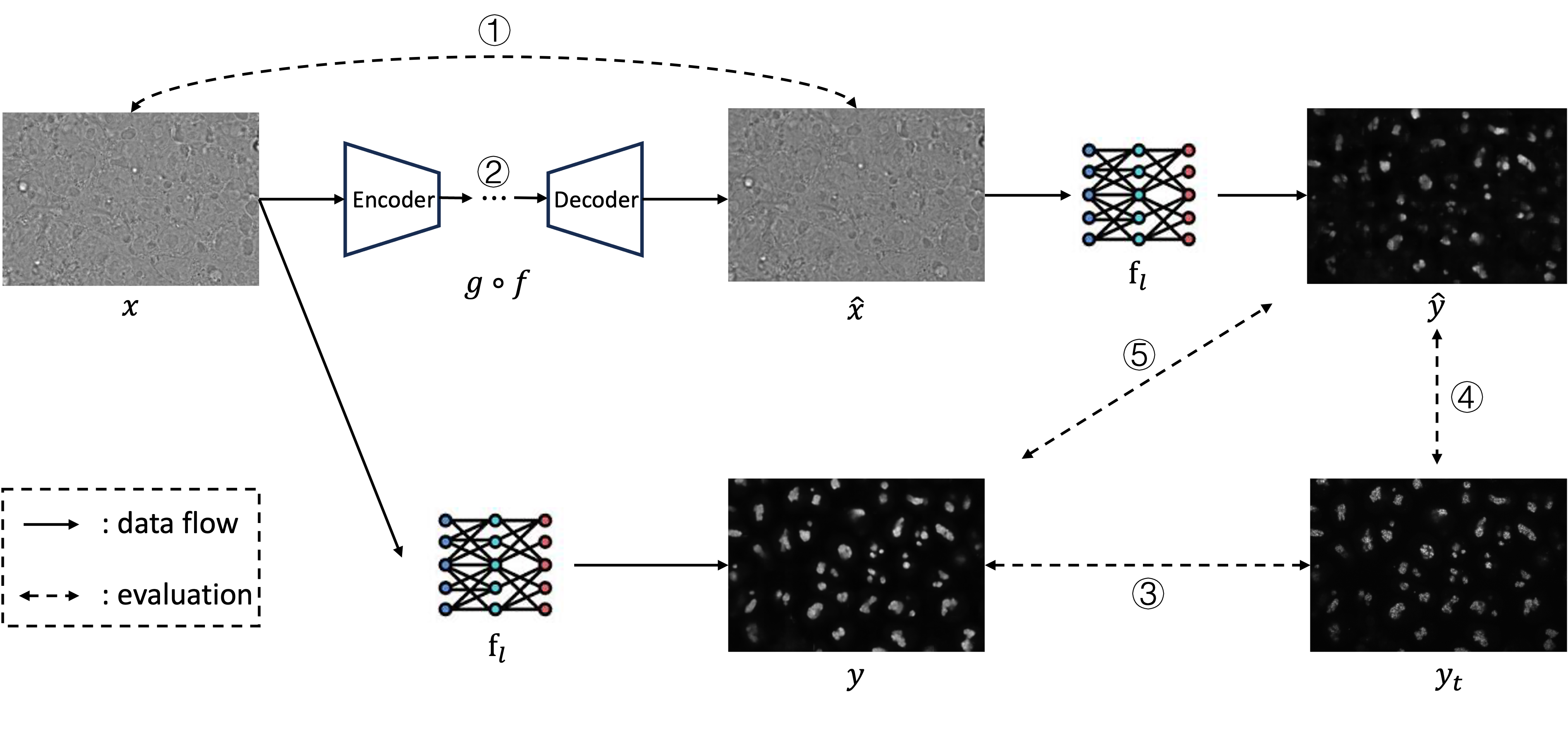}
        \caption{Overview of our proposed evaluation pipeline. The objective is to fully estimate the compression performance of different compression algorithms (denoted as $g \circ f$) in the bioimage field and investigate their influence to the downstream AI-based bioimage analysis tasks (e.g. label-free task in this study, denoted as $f_l$). The solid line represents data flow while the dash line means evaluation. The brightfield raw image $x$ will be compressed and decompressed: $\hat{x} = (g \circ f)(x) = g(f(x))$. Then we feed the reconstructed $\hat{x}$ to the label-free model $f_l$ to get the estimated fluorescent image $\hat{y}$: $\hat{y} = f_l(\hat{x})$. Meanwhile, normal prediction $y$ is also made by $f_l$ from the raw image $x$: $y = f_l(x)$. Regarding the evaluation, \ding{172}\textbackslash\ding{173} exhibits the Rate-Distortion ability of the compression algorithm, \ding{174}\textbackslash\ding{175}\textbackslash\ding{176} represents their influence to the downstream task $f_l$. Specifically, \ding{172} measures the reconstruction ability of the compression method while \ding{173} records the bit-rate and can reflect the compression ratio ability. \ding{174} and \ding{175} represents the prediction accuracy of the $f_l$ model using the raw image $x$ and the reconstructed image $\hat{x}$ as input, respectively. \ding{176} measures the similarity between these two predictions. }
        \label{fig:methodology}
    \end{figure}

The evaluation pipeline was proposed in this study to benchmark the performance of the compression model in the bioimage field and estimate their influence to the downstream label-free generation task. Illustrated in \cref{fig:methodology},  the whole pipeline contains two parts: compression part: $x \xrightarrow{g \circ f} \hat{x}$ and downstream label-free part: $(x / \hat{x}) \xrightarrow{f_l} (y / \hat{y})$, where the former is designed to measure the Rate-Distortion performance of the compression algorithms and the latter aims to quantify their influence to the downstream task. 

During the compression part, the raw image $x$ will be transformed to the reconstructed image $\hat{x}$ through the compression algorithm $g \circ f$:
\begin{equation}
    \hat{x} = (g \circ f)(x) = g(f(x))
\end{equation}
where $f$ represents the compression process and $g$ denotes the decompression process. Note that the compression methods could be both classic strategies (e.g. jpeg) and deep-learning based algorithms. The performance of the algorithm can be evaluated through Rate-Distortion performance, as explained in \cref{eq:R,eq:D,eq:L}. 

In the downstream label-free part, the prediction will be made by the model $f_l$ using both the raw image $x$ and the reconstructed image $\hat{x}$:
\begin{equation}
    y = f_l(x), \hat{y} = f_l(\hat{x}) \label{eq:labelfree}
\end{equation}
The evaluation to measure the compression influence to the downstream tasks is made by:
\begin{gather}
    V = [ y,\hat{y},y_t ] \quad,\quad \rho = [\rho_i]_{i=1}^{4}\\
    S = \left\{ (V_i, V_j) \mid i,j \in \{1,2,3\}, i \neq j \right\} \\
    L = \left[ \rho(V_i, V_j) \mid (V_i, V_j) \in S \right]
\end{gather}
where the evaluation metric L is the collection of different metrics $\rho_i$ on different image pairs $S_i$. V is the collection of the raw prediction $y$, prediction made by the reconstructed image $\hat{y}$ and the ground truth $y_t$. S is formed by pairwise combinations of elements from V. $\rho$ represents the metric we used to measure the relation between image pairs. In this study we totally utilized four metrics: Mean Squared Error (MSE), Structural Similarity Index Measure (SSIM), Peak Signal-to-Noise Ratio (PSNR) and Pearson Correlation.

To conclude, through the above proposed two-phase evaluation pipeline, the compression performance of the compression algorithm will be fully estimated and their impact to the downstream task will also be well investigated.
%--------------------------------------------------------
%--------------------------------------------------------

\section{Experimental settings}
\label{sec:implementation_detail}
\subsection{Dataset}
The dataset used in this study is the human induced Pluripotent Stem Cells (hiPSC) single cell image dataset \cite{forCellScience2018HiPSCDataset} released by the Allen Institute for Cell Science. We utilized grayscale brightfield images and its corresponding fluorescent image pairs from the fibrillarin (FBL) cell line, where the dense fibrillar component of nucleolus is endogenously tagged. For 3D experiments, 500 samples were chosen from the dataset with 395 for training and the remaining 105 samples for evaluation. While in terms of 2D experiments, the middle slice of each 3D sample was extracted, resulting in 2D slices of 624 $\times$ 924 pixels.

\subsection{Implementation Details}
During the first compression part of the proposed two-phase evaluation pipeline, we made the comparison using both classic methods and deep-learning based algorithms. In terms of the classic compression, we employed the Python package 'tifffile' to apply 3 classic image compression: JPEG 2000, JPEG XR, LERC, focusing on level 8 for the highest image quality preservation. To enhance compression efficiency, we used a 16$\times$16-pixel tile-based approach, facilitating image data access during compression and decompression. This methodology enabled a thorough exploration of the storage vs. image quality trade-off.

Regarding learned-based methods, 6 pre-trained models proposed in \cite{Balle2018VariationalHyperprior, Minnen2018JointCompression, Cheng2020LearnedModules} were applied in 2D compression, with each kind of model trained with 2 different metrics (MSE and MS-SSIM), resulting in 12 models in total. The pretrained checkpoints were provided by the CompressAI tool \cite{begaint2020compressai}.  For the 3D senario, an adapted bmshj2018-factorized compression model \cite{Balle2018VariationalHyperprior} was trained and evaluated on our microscopy dataset. For the first 50 epochs, MSE metric was employed in the reconstruction loss term, followed by MS-SSIM metric for another 50 epochs to enhance the image quality.  

When it comes to the second label-free generation part, the pretrained Pix2Pix 2D (Fnet 2D as the generator) and Fnet 3D model were obtained from the mmv\_im2im Python package \cite{SonneckMMV_Im2Im:Transformation}. All the labelfree 2D models were trained by images compressed with the JPEGXR algorithm. For 3D models, raw images were involved in the training.

\section{Results and Discussion}
\label{sec:result_and_discussion}
In this section, we will present and analyze the performance of the image compression algorithms and their impact on the downstream label-free task, using the proposed two-phase evaluation pipeline.

\subsection{Data Compression Results}
\label{subsec:compression_result}
Firstly, we did the compression performance comparison experiment in the context of grayscale microscopic bright-field image, based on the first part of the evaluation pipeline. The results show that deep-learning based compression algorithms behave well in terms of the reconstruction quality and compression ratio ability in both 2D and 3D cases, and outperform the classic methods. 

% Supplementary \cref{tab:metrics_input,tab:compression_ratio} illustrated the quantitative rate-distortion performance of the compression algorithms in the context of 2D grayscale microscopic bright-field images, and a sample result is visualized in \cref{fig:plot_bmshj2018-factorized_ms-ssim_8}. 

The second to the fourth rows in \cref{tab:metrics_input,tab:compression_ratio} demonstrate the quantitative rate-distortion performance for the three traditional compression techniques involved. Although the classic method LERC achieved highest result in all the quality metrics for the reconstructed image, it just saves 12.36\% of the space, which is way lower compared to the deep learning-based methods. Meanwhile, JPEG-2000-LOSSY can achieve comparable compression ratio with respect to AI-based algorithms, but its quality metric ranks the bottom, with only 0.1576 in correlation and 0.4244 in SSIM. The above results compellingly showcase that the classic methods cannot make a trade-off in the rate-distortion performance. 

\begin{figure}[ht]
    \centering
    \includegraphics[width=\textwidth]{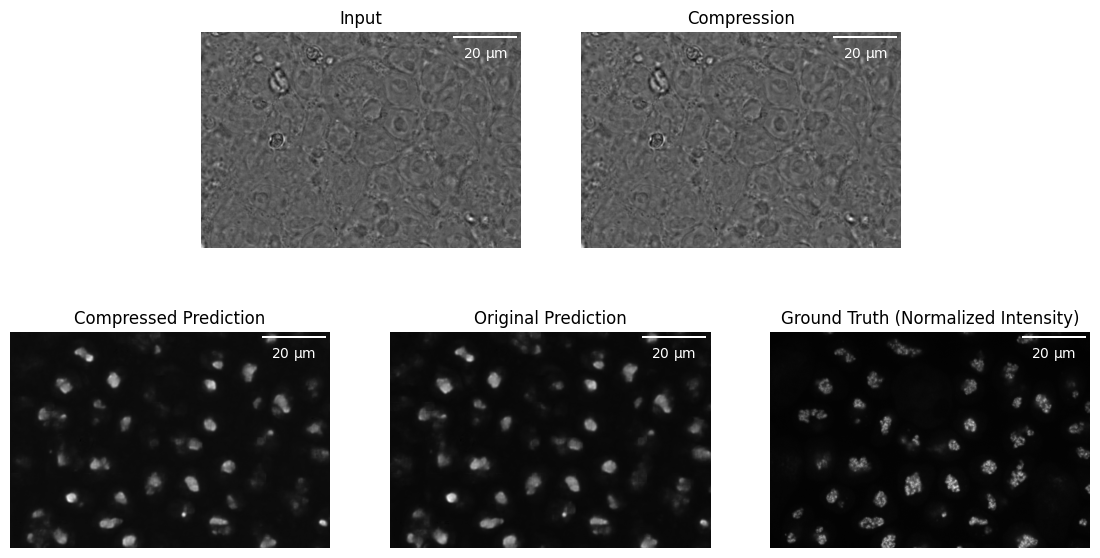}
    \caption{Visualization of 2D brightfield image compression  result (first row, model: bmshj2018-factorized (MS-SSIM)) + downstream label-free model prediction (second row). The upper right compression result is visually plausible compared to the input, and the compressed prediction (bottom left) using the label-free model is very close to the original prediction (bottom middle), which suggests the minimal influence of the selected deep-learning based compression to the downstream task.}
    \label{fig:plot_bmshj2018-factorized_ms-ssim_8}
\end{figure}

Besides, results from deep learning models exhibit close similarities, yielding favorable outcomes, as illustrated in \cref{tab:metrics_input,tab:compression_ratio} from the fifth row to the last. Notably, the 'mbt2018-ms-ssim-8' method exhibits a slight advantage in terms of SSIM, achieving a value of 0.9705. Conversely, the `mbt2018-mean-ms-ssim-8' method showcases a slight edge in correlation, with a score of 0.9866. When considering compression ratio, 'cheng2020-anchor-mse-6' outperforms the others, with an compression ratio of 47.2978. A sample result is visualized in \cref{fig:plot_bmshj2018-factorized_ms-ssim_8}.

\begin{table}[tbp]
  \centering
  \caption{Evaluation of the average 2D bright-field image quality for the different compression methods compared to the original image, to test the reconstruction ability. First column: compression methods, with the second to the fourth rows as the classic methods and fifth to the last as the deep-learning based methods. The second to the last columns indicate the four metrics that we use to measure the reconstruction ability: MSE (the smaller the better), SSIM, Correlation, PSNR (the larger the better)}
  \label{tab:metrics_input}
  \resizebox{\linewidth}{!}{
  \begin{tabular}{l|cccc}
    \toprule
    
    \textbf{Compression} & \textbf{MSE ($px^2$)} & \textbf{SSIM} & \textbf{Correlation} & \textbf{PSNR ($dB$)} \\
    \midrule
    original & 0 & 1 & 1 & 108.1308 \\
    \midrule
    JPEGXR & 0.0011 & 0.8284 & 0.899 & 30.4992 \\
    JPEG-2000-LOSSY & 0.0498 & 0.4244 & 0.1576 & 15.8519 \\
    LERC & 0.0061 & \textbf{0.9803} & \textbf{0.9934} & \textbf{51.72} \\
    \midrule
    bmshj2018-factorized-mse-8 & 0.0002 & 0.9623 & 0.9844 & \textbf{38.4742} \\
    bmshj2018-factorized-ms-ssim-8 & 0.0003 & 0.9704 & 0.9859 & 37.3699 \\
    bmshj2018-hyperprior-mse-8 & 0.0002 & 0.9585 & 0.9829 & 38.4364 \\
    bmshj2018-hyperprior-ms-ssim-8 & 0.0003 & 0.9691 & 0.9854 & 37.1714 \\
    mbt2018-mean-mse-8 & 0.0002 & 0.9559 & 0.9817 & 37.9746 \\
    mbt2018-mean-ms-ssim-8 & 0.0003 & 0.9704 & \textbf{0.9866} & 37.6723 \\
    mbt2018-mse-8 & 0.0002 & 0.9563 & 0.9823 & 38.1687 \\
    mbt2018-ms-ssim-8 & 0.0003 & \textbf{0.9705} & 0.986 & 37.2873 \\
    cheng2020-anchor-mse-6 & 0.0006 & 0.9133 & 0.961 & 34.3726 \\
    cheng2020-anchor-ms-ssim-6 & 0.0009 & 0.9537 & 0.9738 & 33.4254 \\
    cheng2020-attn-mse-6 & 0.0005 & 0.9138 & 0.9609 & 34.5608 \\
    cheng2020-attn-ms-ssim-6 & 0.0006 & 0.9538 & 0.9727 & 34.0428 \\
    \bottomrule
  \end{tabular}
  }
\end{table}

%--------------------------------------------------------
\begin{table}[htbp]
\centering
\caption {Evaluation of the compression ratio and space saving for the different compression methods. First column: compression methods, with the second to the fourth rows as the classic methods and fifth to the last as the deep-learning based methods. The 'Compression ratio' column calculates the ratio between the theoretical image size and the size of the stored bitstream file (the larger the better), while the 'space savings' column is derived from one minus the reciprocal of the previous column. (the larger the better)} \label{tab:compression_ratio}
\resizebox{\linewidth}{!}{
    \begin{tabular}{l|cc}%
        \toprule
        \bfseries Compression & \bfseries Compression ratio & \bfseries Space saving ($\%$)\\
        \midrule
        original & 1.1236 & 10.94 \\
        \midrule
        JPEGXR & 1.3458 & 24.58 \\
        JPEG-2000-LOSSY & \textbf{28.5981} & \textbf{93.70} \\
        LERC & 1.1419 & 12.36 \\
        \midrule
        bmshj2018-factorized-mse-8 & 15.9426 & 93.64 \\
        bmshj2018-factorized-ms-ssim-8 & 19.3469 & 94.82 \\
        bmshj2018-hyperprior-mse-8 & 21.3869 & 95.07 \\
        bmshj2018-hyperprior-ms-ssim-8 & 23.2744 & 95.68 \\
        mbt2018-mean-mse-8 & 23.4083 & 95.50 \\
        mbt2018-mean-ms-ssim-8 & 23.1368 & 95.65 \\
        mbt2018-mse-8 & 23.746 & 95.55 \\
        mbt2018-ms-ssim-8 & 22.9054 & 95.61 \\
        cheng2020-anchor-mse-6 & \textbf{47.2978} & \textbf{97.81} \\
        cheng2020-anchor-ms-ssim-6 & 38.0068 & 97.35 \\
        cheng2020-attn-mse-6 & 47.0159 & 97.80 \\
        cheng2020-attn-ms-ssim-6 & 37.3312 & 97.30 \\
        \bottomrule
    \end{tabular}
}
\end{table}

%--------------------------------------------------------
% \begin{figure}[!htbp]
%     \centering
%     \begin{subfigure}[h]{0.75\textwidth}
%         \centering
%         \includegraphics[width=\linewidth]{figure/JPEGXR_JPEGXR_pred_scale.png}
%         \caption{Prediction from a label-free model trained with JPEG XR compressed images.}
%         \label{fig:ex1_pred}
%     \end{subfigure}
%     \quad
%     \begin{subfigure}[h]{0.75\textwidth}
%         \centering
%         \includegraphics[width=\textwidth]{figure/arrows_scale.png}
%         \caption{Prediction from a label-free model trained with losslessly compressed images \cite{SollmannFOMPrediction}.}
%         \label{fig:ex1_pred_artifacts}
%     \end{subfigure}
%     \caption{The label-free model trained on uncompressed data fails to produce accurate results when applied to lossy compressed images, as evidenced by the visible artifacts. This highlights the incompatibility between the model trained on original data and the application of lossy compression.}
%     \label{fig:result_classic}
% \end{figure}
\begin{figure}[!htbp]
    \centering
    \includegraphics[width=\linewidth]{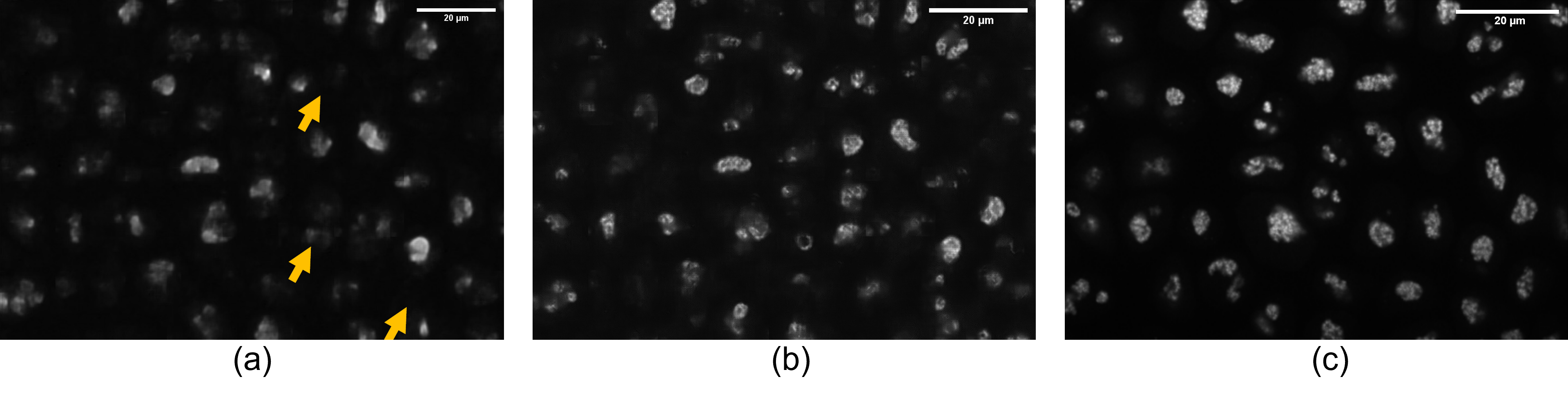}
    \caption{The prediction result of the downstream label-free models trained with lossy/losslessly compressed images, respectively. The input is the lossy compressed bright-field images. (a) Prediction from a label-free model trained with losslessly compressed images \cite{SollmannFOMPrediction}, (b) Prediction from a label-free model trained with JPEG XR compressed images, (c) The ground truth. The label-free model trained on uncompressed data fails to produce accurate results when applied to lossy compressed images, as evidenced by the visible artifacts. This highlights the incompatibility between the model trained on original data and the application of lossy compression.}
    \label{fig:result_classic}
\end{figure}
%--------------------------------------------------------

 Illustrated in \cref{fig:result_3d}, the 3D compression result is visually plausible and the quantitative evaluation metrics are listed in the first row in \cref{tab:metrics_3d}. The metrics are relatively high, reaching 0.922 in SSIM and 0.949 in correlation.  Regarding the compression ratio, 97.74 $\%$ of space will be saved.

In brief, the above findings suggest that deep-learning based compression methods behave well in the context of microscopic image field, averagely outperform the classic methods in terms of reconstruction ability and compression ratios.
 % These findings suggest that although the compressed image reaches relatively high metrics, the following bioimage analysis task will still be affected. The 3D experiments on other learn-based compression methods will be added soon.

 \subsection{Downstream Label-free Results}
\label{subsec:label-free_result}
We also conducted an experiment to assess the impact of the aforementioned compression techniques on downstream AI-based bioimage analysis tasks, specifically the label-free task in our study. Our results indicate that in 2D cases, the prediction accuracy is higher when the input image is compressed using deep learning-based methods, as opposed to traditional methods. Furthermore, this accuracy closely aligns with the predictions derived from the raw image, suggesting that deep-learning based compression methods have a minimal impact on the downstream task.

\begin{figure}[!htbp]
        \centering
        \includegraphics[width=\linewidth]{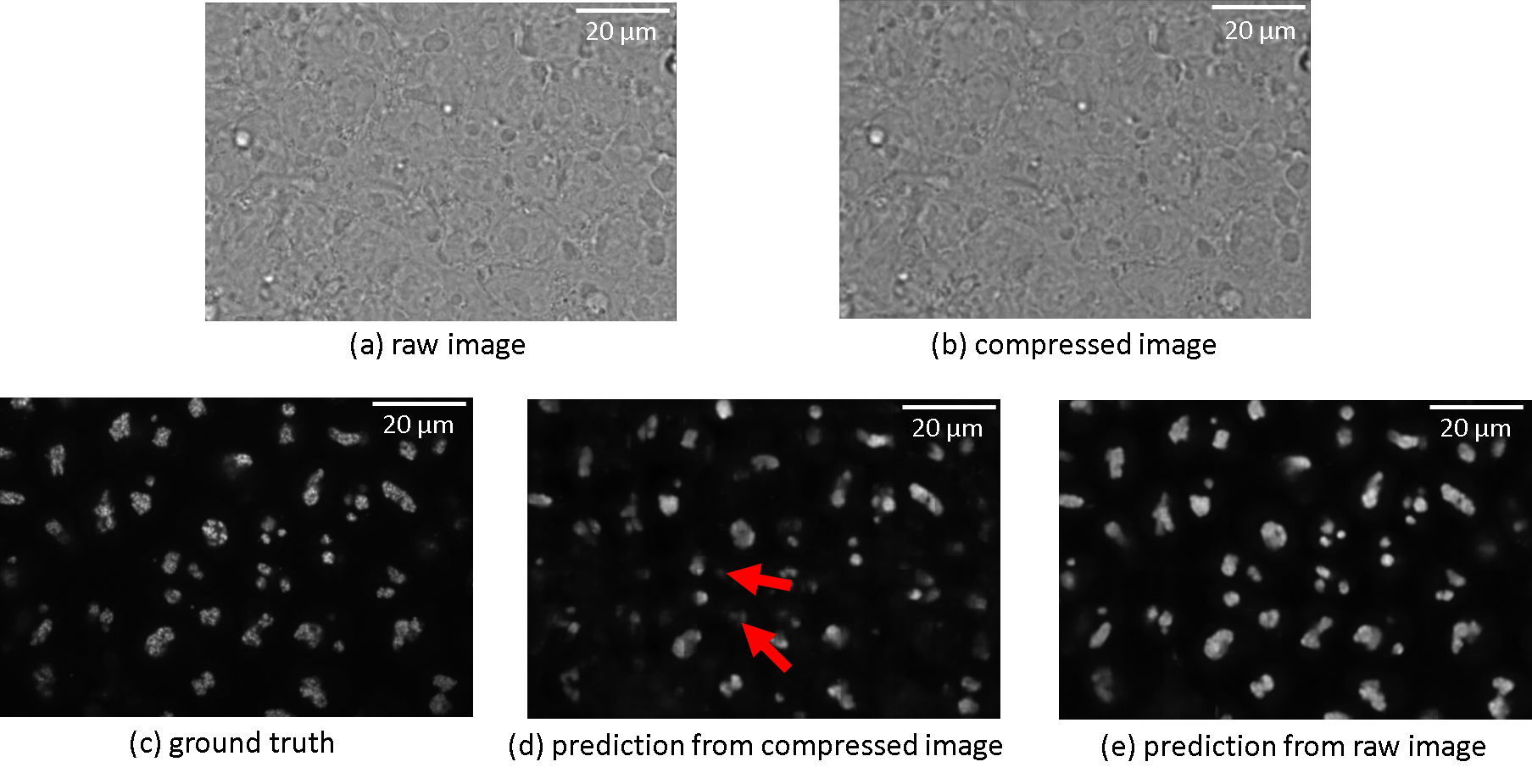}
        \caption{Visualization of 3D compression result based on the bmshj2018-factorized model}
        \label{fig:result_3d}
    \end{figure}

 \cref{tab:metrics_prediction,tab:metrics_prediction_original} exhibits the influence of data compression to the downstream label-free task in 2D cases. Regarding the comparison of the accuracy between the predictions using compressed input and original input (\cref{tab:metrics_prediction}), we found that although the slight degradation in correlation and PSNR, the average SSIM value among deep learning-based methods is akin to the original prediction and surpasses the classic methods, with 'mbt2018-ms-ssim-8' model reaching the highest value (0.7439). If we compare the similarity between the predictions using compressed images and original images (\cref{tab:metrics_prediction_original}), 'mbt2018-ms-ssim-8' and LERC ranked the highest in SSIM and correlation, respectively.
%--------------------------------------------------------
 \begin{table}[!htbp]
  \centering
  \caption{Evaluation of the average prediction quality for the different compression methods compared to the ground truth, to test the impact of the compression methods to the label-free task. First column: compression methods, with the second to the fourth rows as the classic methods and fifth to the last as the deep-learning based methods. }
  \label{tab:metrics_prediction}
  \resizebox{\linewidth}{!}{
  \begin{tabular}{l|cccc}
    \toprule
    \textbf{Compression} & \textbf{MSE ($px^2$)} & \textbf{SSIM} & \textbf{Correlation} & \textbf{PSNR ($dB$)} \\
    \midrule
    original & 0.005 & 0.7333 & 0.7399 & 23.2962 \\
    \midrule
    JPEGXR & 0.006 & 0.6541 & 0.4304 & 22.4266 \\
    JPEG-2000-LOSSY & 0.0347 & 0.3838 & 0.0238 & 17.3313 \\
    LERC & 0.0056 & \textbf{0.6841} & \textbf{0.7394} & \textbf{22.8782} \\
    \midrule
    bmshj2018-factorized-mse-8 & 0.0044 & 0.7258 & 0.6382 & 23.9536 \\
    bmshj2018-factorized-ms-ssim-8 & 0.0046 & 0.7342 & 0.7093 & 23.7132 \\
    bmshj2018-hyperprior-mse-8 & 0.0047 & 0.7128 & 0.6045 & 23.8299 \\
    bmshj2018-hyperprior-ms-ssim-8 & 0.0043 & 0.7378 & 0.6912 & 23.9392 \\
    mbt2018-mean-mse-8 & 0.0044 & 0.7229 & 0.6097 & 23.9387 \\
    mbt2018-mean-ms-ssim-8 & 0.0044 & 0.7415 & 0.7073 & 23.8704 \\
    mbt2018-mse-8 & 0.0043 & 0.7194 & 0.6207 & \textbf{23.9625} \\
    mbt2018-ms-ssim-8 & 0.0044 & \textbf{0.7439} & \textbf{0.7102} & 23.8387 \\
    cheng2020-anchor-mse-6 & 0.0065 & 0.5991 & 0.4375 & 22.4682 \\
    cheng2020-anchor-ms-ssim-6 & 0.0045 & 0.7145 & 0.6389 & 23.7949 \\
    cheng2020-attn-mse-6 & 0.008 & 0.568 & 0.4208 & 21.8606 \\
    cheng2020-attn-ms-ssim-6 & 0.0044 & 0.718 & 0.6418 & 23.8619 \\
    \bottomrule
  \end{tabular}
  }
\end{table}
%--------------------------------------------------------
%--------------------------------------------------------
\begin{table}[!htbp]
\centering
\caption {Evaluation of the average prediction quality for the different compression methods compared to the original prediction. First column: compression methods, with the second to the fourth rows as the classic methods and fifth to the last as the deep-learning based methods.} \label{tab:metrics_prediction_original}
\resizebox{\linewidth}{!}{
    \begin{tabular}{l|cccc}
    \toprule
    \textbf{Compression} & \textbf{MSE ($px^2$)} & \textbf{SSIM} & \textbf{Correlation} & \textbf{PSNR ($dB$)} \\
    \midrule
    original & 0.0 & 1.0 & 1.0 & 108.1308 \\
    \midrule
    JPEGXR & 0.0061 & 0.8255 & 0.573 & 22.4105 \\
    JPEG-2000-LOSSY & 0.0323 & 0.4835 & 0.0346 & 16.7578 \\
    LERC & 0.0003 & 0.9624 & 0.9995 & 52.1393 \\
    \midrule
    bmshj2018-factorized-mse-8 & 0.0026 & 0.9103 & 0.859 & 27.0981 \\
    bmshj2018-factorized-ms-ssim-8 & 0.0009 & 0.9541 & \textbf{0.9483} & 30.9527 \\
    bmshj2018-hyperprior-mse-8 & 0.0034 & 0.8904 & 0.8158 & 25.9064 \\
    bmshj2018-hyperprior-ms-ssim-8 & 0.0014 & 0.9427 & 0.9232 & 29.4052 \\
    mbt2018-mean-mse-8 & 0.0031 & 0.8983 & 0.8161 & 26.1503 \\
    mbt2018-mean-ms-ssim-8 & 0.001 & 0.9516 & 0.9451 & 30.7766 \\
    mbt2018-mse-8 & 0.003 & 0.8989 & 0.8327 & 26.4424 \\
    mbt2018-ms-ssim-8 & 0.0009 & \textbf{0.9551} & \textbf{0.9483} & \textbf{31.0031} \\
    cheng2020-anchor-mse-6 & 0.0066 & 0.7776 & 0.5942 & 22.2884 \\
    cheng2020-anchor-ms-ssim-6 & 0.0026 & 0.9078 & 0.8499 & 26.3781 \\
    cheng2020-attn-mse-6 & 0.0078 & 0.7452 & 0.5713 & 21.769 \\
    cheng2020-attn-ms-ssim-6 & 0.0026 & 0.9098 & 0.8523 & 26.5122 \\
    \bottomrule
  \end{tabular}
}
\end{table}
%--------------------------------------------------------
When it comes to 3D cases, The prediction from the compressed image is not comparable to that predicted by the raw bright-field image (2.54 dB $\downarrow$ in PSNR and 0.08 $\downarrow$ in SSIM), as shown in the second and third rows from \cref{tab:metrics_3d}, indicating a quality downgrade during compression. This can be attributed primarily to the ignorance of considering compression in the training phase of the labelfree model, which will be discussed subsequently. Illustrated in \cref{fig:result_3d}, despite the visually plausible reconstruction result, the information loss during the compression process also heavily affects the downstream label-free generation task. For instance, The fibrillarin structure pointed by the arrow in the prediction result from the compressed image is missing, which is quite obvious in the corresponding prediction from the raw image.

Briefly, the above result suggests that in 2D cases, the downstream task will be less affected when deep-learning based methods were applied. However, the prediction accuracy will be largely affected in 3D cases.

%--------------------------------------------------------
\begin{table}[htbp]
    \centering
    \caption {3D compression results based on the bmshj2018-factorized model. Both compression performance (the first row) and effect on the downstream task (2-4 rows) are evaluated.}
    \label{tab:metrics_3d}
    \resizebox{\linewidth}{!}{
    \begin{threeparttable}
        \begin{tabular}{c|cccc}%
            \toprule
            \bfseries Comparison* & \bfseries MSE & \bfseries SSIM & \bfseries Correlation & \bfseries PSNR ($dB$)\\
            \midrule
            (a) to (b) & 0.3344 & 0.9220 & 0.9484 & 28.1366 \\
            (d) to (c) & 0.0010 & 0.8495 & 0.5981 & 30.0605 \\
            (e) to (c) & 0.0006 & 0.9268 & 0.9066 & 32.6057 \\
            (d) to (e) & 0.0015 & 0.8203 & 0.6576 & 28.5425 \\
            \bottomrule
        \end{tabular}
        \begin{tablenotes}
            \item[] \footnotesize *The index here is consistent with \cref{fig:result_3d}. (a). raw bright-field image; (b). compressed bright-field image; (c). label-free ground truth (d). label-free prediction from compressed image; (e). label-free prediction from raw image
        \end{tablenotes}
        \end{threeparttable}
    }
\end{table}
%--------------------------------------------------------
 % Based on the above results in \cref{subsec:compression_result,subsec:label-free_result}, we can intuitively discern that when the reconstructed images exhibit higher quality metrics (indicating a closer resemblance to the original images), they generally have a smaller impact on downstream tasks, and vice versa. Since the downstream tasks are trained using the uncompressed original images, we can readily deduce the following proposition: within a certain degree of generalization capacity, the performance of downstream models is contingent on the distribution disparity between input images and training images. To validate this proposition, we devised the following experiment: two label-free models using uncompressed data and JPEGXR-compressed data as input were trained respectively and we compared the performance of these models on JPEGXR compressed input images. Illustrated in \cref{fig:result_classic},  we observed significant artifacts in the prediction when the model was not trained on the compressed data used as input, which is subject to the low quality metrics shown in \cref{tab:metrics_input}. However, artifacts were almost mitigated when the model was trained with data using the same compression algorithm, which has the closer data distribution. The above phenomenon highlights the importance of considering compression in the training process in order to achieve favorable outcomes.
  Given that the 2D labelfree models were all trained with compressed images, it is also crucial to measure the impact of compression during the training phase in the downstream labelfree task. For this purpose, we devised the following experiment: two label-free models using uncompressed data and JPEGXR-compressed data as input were trained respectively and we compared the performance of these models on JPEG XR compressed input images. Illustrated in \cref{fig:result_classic},  we observed significant artifacts in the prediction when the model was not trained on the compressed data used as input, which is subject to the low quality metrics shown in \cref{tab:metrics_input}. However, artifacts were almost mitigated when the model was trained with data using the same compression algorithm, which has the closer data distribution. This may also explains the precision drop in 3D cases, where the compression were not considered in the training phase. The above phenomenon highlights the importance of considering compression in the training process in order to achieve favorable outcomes.

%--------------------------------------------------------
    
\section{Conclusion}
\label{sec:conclusion}
In this research, we proposed a two-phase evaluation pipeline, in order to benchmark the rate-distortion performance of different data compression techniques in the context of grayscale microscopic brightfield images and fully explored the influence of such compression to the downstream label-free task. We found that AI-based image compression methods can significantly outperform classic compression methods and have minor influence on the following label-free model prediction. Despite some limitations, we hope our work can raise the awareness of the application of deep learning-based image compression in the bioimaging field and provide insights into the way of integration with other AI-based image analysis tasks.

\bibliography{data_compression}% common bib file        

\end{document}